\documentclass[5p]{elsarticle}

\usepackage{lineno,hyperref}
\modulolinenumbers[5]
\usepackage{graphicx} 
\usepackage{caption}
\usepackage{subfig}
\usepackage{multirow}
\usepackage{float} 
\usepackage{color}

\journal{Engineering Applications of Artificial Intelligence}  

\begin{document}

\begin{frontmatter}

\title{Tackling the muon identification in water Cherenkov detectors problem for the future Southern Wide-field Gamma-ray Observatory by means of Machine Learning}



\author[lisboa,granada]{B.S. Gonz\'alez}
\author[lisboa]{R. Concei\c{c}\~ao}
\author[lisboa]{M. Pimenta}
\author[lisboa]{B. Tom\'e}
\author[granada]{A. Guill\'en}

\address[lisboa]{Laborat\'orio de Instrumenta\c{c}\~ao e F\'isica Experimental de Part\'iculas (LIP) - Lisbon,
Av. Prof. Gama Pinto 2, 1649-003 Lisbon, Portugal and
Instituto Superior T\'ecnico (IST), Universidade de Lisboa, Av. Rovisco Pais 1, 1049-001 Lisbon, Portugal
}
\address[granada]{Computer Architecture and Technology Department, University of Granada, Granada, Spain}

\begin{abstract}
This paper presents several approaches to deal with the problem of identifying muons in a water Cherenkov detector with a reduced water volume and 4 PMTs. Different perspectives of information representation are used and new features are engineered using the specific domain knowledge. As results show, these new features, in combination with the convolutional layers, are able to achieve a good performance avoiding overfitting and being able to generalise properly for the test set. The results also prove that the combination of state-of-the-art Machine Learning analysis techniques and water Cherenkov detectors with low water depth can be used to efficiently identify muons, which may lead to huge investment savings due to the reduction of the amount of water needed at high altitudes. This achievement can be used in further research to be able to discriminate between gamma and hadron induced showers using muons as discriminant. 
\end{abstract}


\end{frontmatter}


\section{Introduction}

The irruption of Deep Learning (DL) models using Convolutional Neural Networks (CNNs) has been earthshaking for the research in machine learning by setting the standard outperforming human capabilities in many tasks. It all started in image applications with \cite{lecun1989handwritten} but, rapidly, these methods have been applied to many other domains \cite{KIRANYAZ2021107398,ERHAN202164}. 
Outside the image landscape, the principal application of these models has been biomedical signals like ElectroCardioGrams (ECG) and ElectroEncephaloGrams (EEG) \cite{Peimankar2021,ManzanoGRH17, KIRANYAZ2021107398}.  

For the concrete case of analysing cosmic rays measurements, DL has been successfully applied to a few experiments. Some of them resemble the application done in image processing as images can be generated in a quite straight-forward way from the experimental devices as in Imaging Atmospheric Cherenkov Telescope Arrays (IACTAs). However, a different approach is required to process other type of inputs in other detectors,
for instance, the reconstruction of the main characteristics of Extensive Air Showers (EAS) in observatories like HAWC or Pierre Auger \cite{GUILLEN201912,Guillen2020, capistran2015new} or the detection of neutrinos with the IceCube observatory \cite{choma2018graph}. 


In this research field, the detection of very-high-energy (VHE) gamma-rays is essential to investigate the sources of the incoming electromagnetic radiation produced by some of the most extreme, non-thermal, phenomena taking place in the Universe, e.g., fast-spinning neutron stars or supermassive black holes \cite{Pimenta2018Astroparticle}. One of the possible techniques to indirectly detect gamma-rays are the EAS arrays, which cover large areas with particle detectors at high altitude. This method reconstructs the main characteristics (e.g. energy or direction) of the primary gamma-ray by means of the detection of secondary particles produced during the air shower development.  EAS arrays are able to perform long term observations of variable sources and allow the search for emissions in extended regions \cite{assis2017lattes}.

However, although the indirect methods are effective in the GeV-TeV region, an important disadvantage is the presence of a huge background from charged particles produced in showers originated by cosmic-rays, which is three orders of magnitude larger than the signal. Several techniques have been proposed to perform the gamma/hadron separation, that is, rejecting the hadronic background. 
At high energy, muons, a clear signal of hadronic interactions, begin to reach the ground in sufficiently high numbers so that they can be used to discriminate gamma from hadronically induced showers. Thus, the identification of muons in water Cherenkov detectors (WCD) is explored in this work. 
 
Using light detectors, WCDs measure the Cherenkov light emitted by charge particles in water. 
The production of the Cherenkov light in WCDs is closely related to its depth. Since electromagnetic particles get attenuated at the top of the station, using a detector unit with enough depth ensures a good efficiency in the identification of muons. That is the case of HAWC observatory, which has a good discrimination power in single muons events (those events with only muons and no electromagnetic contamination) \cite{zuniga2017detection,barber2017detection} by using 300 cylindrical WCDs with a depth of 5 meters and a diameter of 7,5 meters. However, it must be taken into account that at mountain altitudes the resources of liquid water are reduced, then, the detector size must be optimised to reduce the necessity of water at those altitudes. For this reason, in this paper, 
the problem is tackled from a new WCD design with low water depth and a good muon tagging efficiency. 
The proposed detector uses a set of four Photomultipliers Tubes (PMTs), whose positions inside the WCD have been optimised to maximise the signal asymmetric of muons vertically crossing the detector. 



The challenge in this work is to tackle the problem of identifying if a muon has crossed the WCD by means of the PMTs' gathered information. Once it is known if there are muons in the event, this information could be used for an ulterior classification between $\gamma$ and hadron induced showers which could be straight applied to new observatories. 
To do so, it is necessary to establish the whole machine learning pipeline as well as determine how the input information will be given to the models. Thus, 
the rest of the paper is organised as follows: in Section \ref{sec:data}, the data for the analysis is described. Section \ref{sec:encoding} presents the different approaches considered to tackle the problem. Section \ref{sec:results} performs an experimental comparison of the previous approaches. Finally, the conclusions and a discussion are presented in Section \ref{sec:conclusions}.

\section{Data description and preprocessing} \label{sec:data}

\begin{figure*}[t]
 \centering
  \subfloat[Single-layered WCD design.]{
   \label{fig:WCD_design}
    \includegraphics[width=0.4\textwidth]{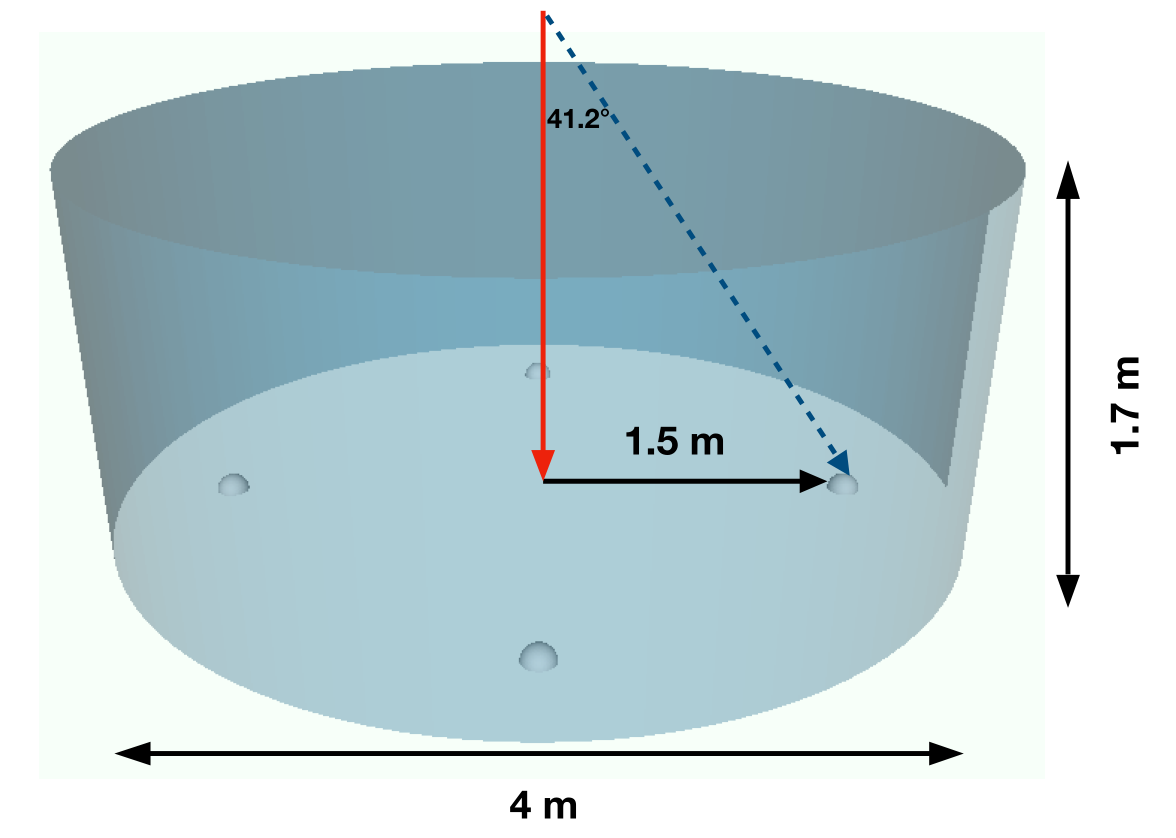}}
\hspace{0.5in}
  \subfloat[Water Cherenkov detector crossed by a \textit{Single Muon}. The WCD drawing is surrounded by the correspondent PMT signal time traces.]{
   \label{fig:WCD_and_trace}
    \includegraphics[width=0.35\textwidth]{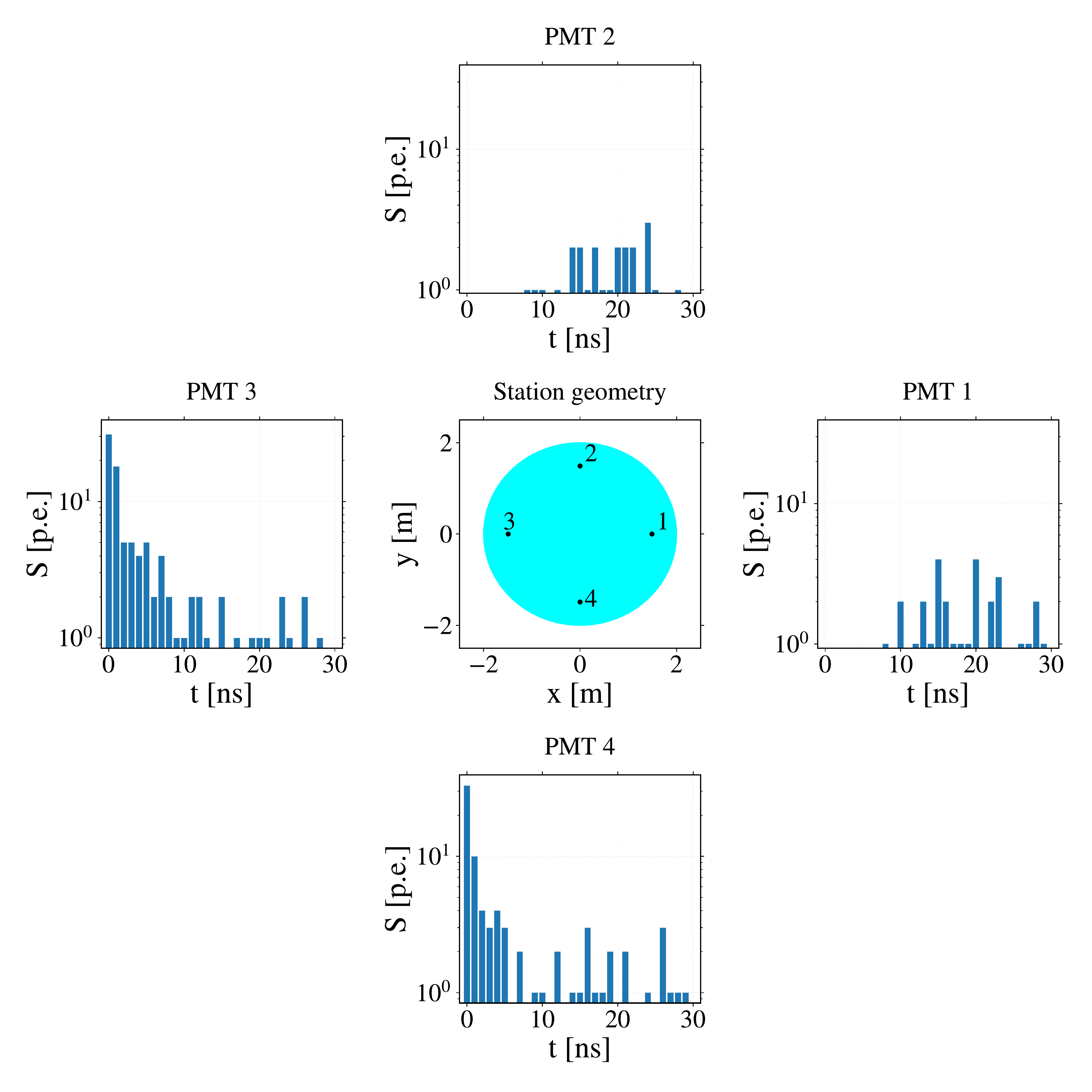}}
 \caption{Scheme of the single-layered WCD.}
 \label{fig:WCD_detector}
\end{figure*}


The data analysed in this work has been simulated using CORSIKA (version 7.5600) \cite{CORSIKA} and the detector using the Geant4 toolkit (version 4.10.05.p01) \cite{agostinelli2003geant4,Geant4_2006,Geant4_2016}. The observation level for the simulations was set at $5\,200\,$ m above the sea level and a WCD array covering an area of $80\,000\,{\rm m^2}$ and a fill factor of $\sim 80\%$ was considered which are the operational characteristics that the future Southern Wide field Gamma-ray Observatory (SWGO) \cite{SWGO} will have.

The set of EAS generated was conducted using proton-based particles with energies between $4 - 6\,$TeV and a zenith angle $\theta_0 \in [5^\circ;15^\circ]$. This range makes muons to fall vertically towards the station, making its presence clearer and adding some variability to the data set instead of having to introduce noise to make them more realistic. The azimuth angle of the primary particle was uniformly distributed over $\phi$. The WCD unit used in this study is a cylinder $1.7\,$m height with a diameter of $4\,$m (see Figure \ref{fig:WCD_design}). The station has $4$ PMTs at the bottom, equally separated and distancing from the WCD center by $1.5\,$m. These were the dimensions encountered to guarantee a good signal uniformity and maximise the signal asymmetry between PMTs for vertically entering muons, as for instance the case shown in Figure \ref{fig:WCD_and_trace}. 

\subsection{Data curation and preprocessing}

A crucial stage in the experimentation is to define a procedure to arrange the data registered by the detectors into a classical machine learning problem. First of all, for each EAS, the simulator considers all the stations that are hit by any particle. However, it is necessary to be cautious and select only the events that are interesting for our purpose. For instance, in the case that a muon doesn't cross the detector entirely, few Cherenkov light could be detected. To overcome possible unfavorable situations a threshold of 300 photoelectrons in the light collected is established when selecting the events. 



The next step is to build the data sets necessary for the experimentation. As the muon identification is carried out at station level, to ensure the maximum fairness in our analysis, the EAS are separated beforehand to avoid having events of the same shower in different data sets. 

Afterwards, the data is partitioned into three independent subsets:

\begin{itemize}
    \item \textbf{Train}: necessary for setting a value for model hyperparameters. 
    \item \textbf{Validation}: used to select the best final model. To this end, a $20 \%$ is taken from the training data set.
    \item \textbf{Test}: applied to perform an unbiased evaluation of the fittest models after the training/validation process. 
\end{itemize}

Table \ref{tab:datasets_description} details the number of instances of each partition. 

\begin{table}[H]
\centering
\scriptsize
\resizebox{7cm}{!} {
\begin{tabular}{c|c|c|}
\cline{2-3}
\multicolumn{1}{l|}{}                                & \multicolumn{2}{c|}{\textbf{Data sets}} \\ \cline{2-3} 
\textbf{}                                            & \textbf{Training  } & \textbf{ Test } \\  \hline
\multicolumn{1}{|c|}{\textbf{EAS}}           & $2 \, 000$                         & $1 \, 637$       \\ \hline
\multicolumn{1}{|c|}{\textbf{S.M. stations}} & $17 \, 244$                        & $14 \, 808$      \\ \hline
\multicolumn{1}{|c|}{\textbf{E.M. stations}}   & $340\,007$                       & $289 \, 354$     \\ \hline
\multicolumn{1}{|c|}{\textbf{Instances}}     & $357 \, 251$                       & $304 \, 162$     \\ \hline
\multicolumn{1}{|c|}{\textbf{Muonic prop.}}  & 4.83 $\%$                    & 4.87 $\%$  \\ \hline
\end{tabular}}
\caption{Description of the data sets. Number of station events with \textit{Single Muons} (S.M. stations) and without muons (E.M. stations)  from proton-induced showers with $E_0 \in [4;6]\,$TeV and $\theta_0 \in [5^\circ;15^\circ]$.}
\label{tab:datasets_description}
\end{table}

Muons constitute a small fraction of the total number of particles that reach the Earth's surface. Thus, as shown in Table \ref{tab:datasets_description}, the number of stations with muons per EAS is low (in comparison with those with electrons or photons). Since the subsets are partitioned by separating the EAS, the proportion of stations with muons will remain. For this reason, different preprocessing techniques have been explored in order to balance the classes before the training stage, which is a must if we want to identify any muon \cite{Gonzalez_2020}.

\begin{itemize}
    \item \textbf{Random oversampling}: random repetition of the samples available from stations with muons.
    \item \textbf{Random undersampling}: random elimination of the samples available from stations without muons.
    \item \textbf{SMOTE} (\textit{Synthetic Minority Over-sampling TEchnique}): creation of synthetic new samples of stations with muons \cite{10.5555/1622407.1622416}.
\end{itemize}


\section{Machine Learning algorithms: problem approach and model design } \label{sec:encoding}
This section presents the different approaches considered to solve the problem as well as the details regarding the design methodology for each type of technique.

\subsection{Problem encoding and feature engineering}


As it is desired to know the probability that the WCD analysed has been crossed by a muon, instead of approaching the task as a classification problem (where output would be \{0,1\}), regression models are the adequate to provide such information. By having the probability, it will be possible to tune the model towards a more sensible or specific output by defining thresholds \cite{guillen2018preliminary}.

The nature of the data allows using different techniques when handling the information collected by the PMTs. In this paper, the following strategies have been considered:

\begin{itemize}
    \item $Signals$ ($\vec{S} = \vec{S_1}, \vec{S_2}, \vec{S_3}, \vec{S_4}$): it corresponds to the signal trace captured of the event during 30 ns, which allows us to explore the temporal features of the events and may be crucial for identifying, in future work, if more than one particle has crossed the station. This is specially interesting considering that analysing the whole signal could make possible to identify reflections of light generated by the tank walls. Each $\vec{S_i}$ is highly dependant on the total energy of the event, therefore, the raw signals are normalised by the sum of the integrals of all PMTs during 30 ns.

    \item PMT Integrals ($I_1, I_2, I_3, I_4$): instead of having such a large amount of data, it is possible to project the traces into a number by computing the integral of the signal recorded by each detector. 
    By doing this, it is possible to consider classical approaches which would not been able to process the raw signals described above. As there are 4 PMTs recording the Cherenkov light, it is still possible to conserve some spatial information although the temporal component is lost. For the sake of clarity in the muon identification, as they reach the PMTs before the electromagnetic component, a threshold of 10 ns has been set to compute these values. 
    
    \item Total light $I_t$: it represents the sum of the PMT integrals, therefore, $I_t = \sum_{i=1}^4 I_i$. The reason to consider $I_t$ is due to the intrinsic uniform properties of muons, which are confined in a concrete range of values of this variable. 
    
    \item Asymmetry $A_T$: Cherenkov light in muonic events is mainly produced in a reduced area of the WCD, whilst in electromagnetic events the light is spread inside the tank. Under these circumstances, it is to be expected that a large asymmetry is present in muonic events when comparing the amount of light collected by the PMT with the maximum signal and the one in front of it. Let $A_{T}$ be the total asymmetry of an event, defined by equation (\ref{eq:asymmetry}), and let $I_{max}$ and $I_{opp}$ be the integral of the signals in the PMT with maximum signal and the one in front of it respectively, so that $A_{T} \in \left[ 0,1 \right]$. 

\begin{equation} \label{eq:asymmetry}
A_{T} = \frac{I_{max} - I_{opp}}{I_{max} + I_{opp}}
\end{equation}

    \item PMT Integrals Normalization ($In_1,...,In_4$): Although the integral by itself could be a good projection of the event, it is possible to engineer the information encoding by normalising the amount of signal of each PMT considering the total signal recorded to obtain a magnitude independent range of values. Thus, the integrals for each PMT are normalised using $I_t$. Let, $In_i = I_i / I_t$.
    
    
    
\end{itemize}





As a summary, Table \ref{tab:variables} shows the input given to the models and its classification.

\begin{table*}[]
    \centering
    \scriptsize
    \begin{tabular}{|c|c|c|c|}
\hline
Variable \# & Variable Name                                                          & Variable Type & Meaning                                                                                                                                                        \\ \hline
1           & $\vec{S}$                                                     & Simulated     & \begin{tabular}[c]{@{}c@{}} PMTs' signal traces. First 30 ns of the signal\\ trace of each PMT. \\ Normalised using the total signal.\end{tabular}                                                                         \\ \hline
2           & $I_t$                                                      & Engineered     & \begin{tabular}[c]{@{}c@{}}WCD's total signal. Sum of the integrals of each \\ PMTs' signal using 10 ns. Total\\ Cherenkov light in the WCD.\end{tabular}                          \\ \hline
3           & \begin{tabular}[c]{@{}c@{}} $I_i, i=1...4$ \end{tabular} & Engineered     & \begin{tabular}[c]{@{}c@{}}Integral of the PMT's signal using\\ the first 10 ns. \end{tabular}            \\ \hline
4           & \begin{tabular}[c]{@{}c@{}} $In_i, i=1...4$ \end{tabular} & Engineered     & \begin{tabular}[c]{@{}c@{}}Integral of the PMT's signal using\\ the first 10 ns. The variable is \\ normalise by the WCD total signal.\end{tabular}            \\ \hline
5           & $A_t$                                                              & Engineered    & \begin{tabular}[c]{@{}c@{}}Asymmetry. A normalised difference of signal \\ between the PMT with the greatest\\ signal and its opposed in the WCD.\\ Defined in equation (\ref{eq:asymmetry}).\end{tabular} \\ \hline
\end{tabular}
    \caption{Variables used in the different approaches. Variable number 3 is actually 4 variables, one for each PMT in the tank. In the same way, variable number 1 consists of 4 $\times$ 30 values corresponding to the trace registered by each of the 4 PMTs.}
    \label{tab:variables}
\end{table*}

\subsection{Model design methodologies}

This subsection first describes a first approach on CNNs application to the problem using the traces available. Afterwards, it is detailed the process followed with boosted tress which are not able to process the raw signals. Finally, in this section, it is commented the possibility of combining those different approaches by means of an ensemble.

\subsubsection{Models using Convolutional Neural Networks}\label{sec:CNN}
To process the signal trace recorded by the PMTs 1-dimensional convolutional neural networks (CNNs) have been designed. As stated previously, CNNs are one of the state-of-the-art machine learning algorithms, whose relevant applications in different fields have proven its great potential. The algorithm is composed of two blocks of hidden-layers: a first set of convolutional layers extract the features from the raw data by means of the convolution, afterwards, a second set of hidden-layers uses the previous information to perform the classification or regression and provide the output. Therefore, CNNs are capable of fusing the tasks of feature extraction from complex data (signals or images) and classification/regression in a single process, which is the main advantage of using this algorithm \cite{kiranyaz20191d}. Before hyperparameters can be tuned, it is required to find a way of introducing such input. As stated previously, different kind of inputs appear after analysing the information captured by the detectors. With the use of this algorithm we intend to squeeze all the information available, both spatial and temporal.

On the one hand, convolutional layers will be used to extract characteristics from the signal traces. For each considered station we will have 4 signal traces to store (one per PMT). 
In order to respect as much as possible the temporal information that the problem provides, the approach consist on using a channel for each of these signals, in this way, each signal trace value will be attached to the nanosecond when it was registered by its position in the array which stores the data. 

On the other hand, after focusing on the temporal characteristics, we must consider how to feed the CNN with the other engineered variables as well. To this end, the second set of hidden layers is fed with the rest of the variables, that is, the dense layers which are used to perform the regression. 

When designing a CNN, one of the critical stages is the definition of the convolutional layers. Nowadays, no standard approach exists to set the configuration of a CNN. There are some rules of thumb for image classification, but, since the problem tackled in this paper belongs to another domain, those rules could not be adequate. To overcome this problem, the hyperparameter space has been clustered based on the experience tackling similar problems like \cite{carrillo2019improving,manzano2017combination}, so that a finite number of values are evaluated. Finally, after evaluating different topologies with the validation data set, the fittest one was composed of three convolutional layers which were enough to cover the inner complexity of the data and extract relevant features from the raw data. ReLU \cite{Goodfellow-et-al-2016} activation function was used in these layers. The size of the filters was selected according to the mean of the signal traces registered, which unveils that the first pulse of Cherenkov light usually arrives within the first 3 nanoseconds of signal. Thus, small filters are proposed. The fittest configuration used filters of size 2, with a stride = 2 in the first layer to highlight the arrival of the direct Cherenkov light. The number of filters in these convolutional layers was 20, 15 and 10. Pooling layers have been discarded since they reduce the average specificity, though higher average sensitivity was achieved when using them. 

Regarding the architecture of the CNN, the configuration chosen is: three dense layers (with 30, 15 and 10 neurons) and a final output layer were used to perform the regression from the previous extracted characteristics and the introduced variables. Sigmoid activation function was used for the three dense layers and for the final neuron. ReLU and Tanh activation functions were considered as well but performance obtained was smaller.

All CNNs configurations (using different input variables) were trained using Adam learning algorithm \cite{kingma2014adam}, which is a stochastic gradient-based method and has been broadly used in many deep learning applications. The adjustment of its parameters was done following those recommended in \cite{kingma2014adam}, that is: betas = (0.9, 0.999) and epsilon = $10^{-7}$ and learning rate of 0.001. The models were trained during 200 epochs with a batch size of 512, smaller values were showing high differences between training and validation and higher values increased training errors. Regarding the loss function to be minimised after each epoch it was chosen Mean Squared Error (MSE). 

\begin{figure}[t]
	\centering
	\includegraphics[width=0.85\linewidth]{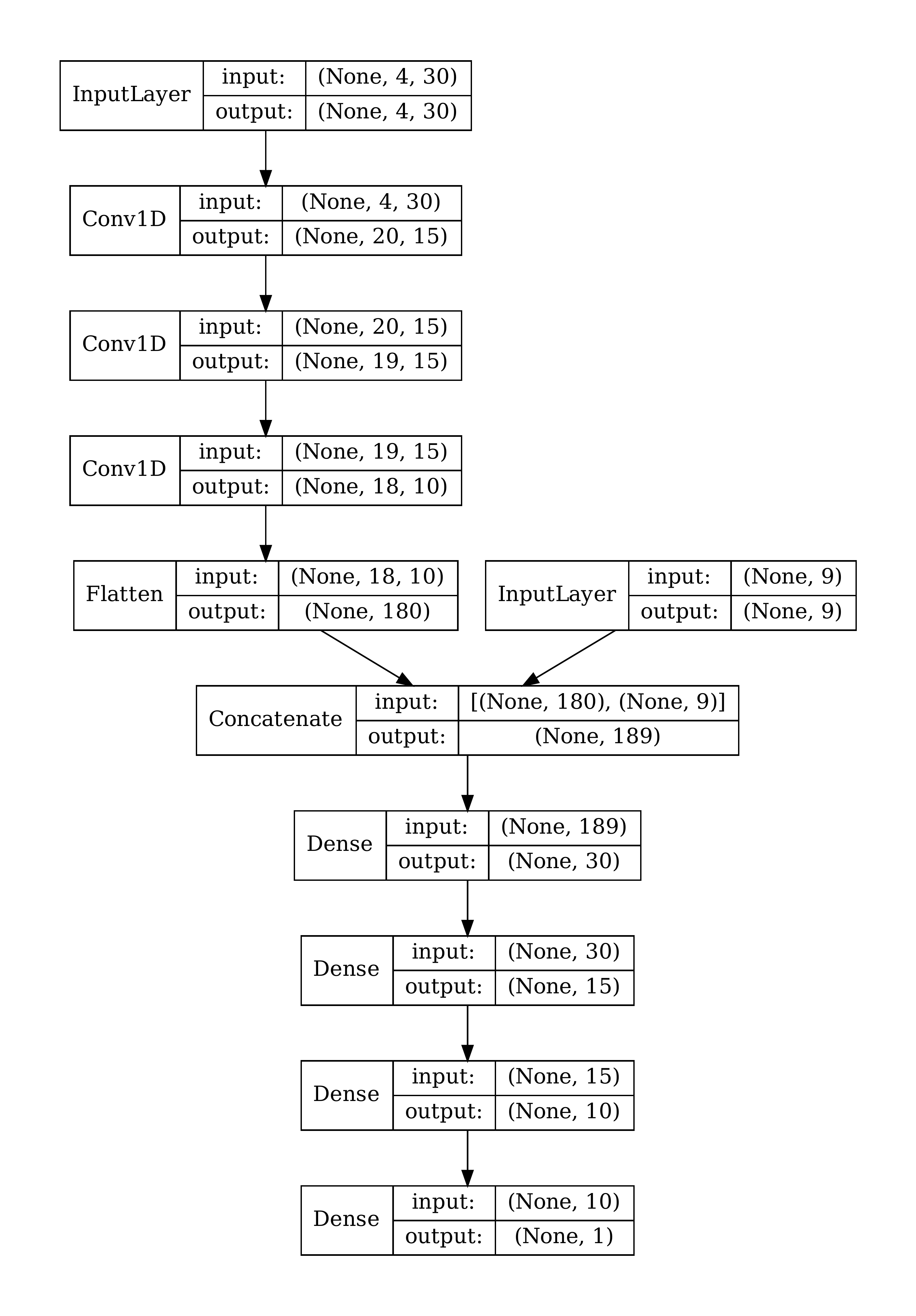}
	\caption{Topology of the fittest CNN found.}
	\label{fig:CNN_topology}
\end{figure}

Figure \ref{fig:CNN_topology} shows the topology of the fittest CNN found using temporal and spatial features, whose parameters have been detailed previously.

\subsection{Models using decision-tree based algorithm}
Two state-of-the-art decision-tree based algorithms have been used to process exclusively the signal traces integrals which enable us to study the spatial information present in the WCDs: \textit{Extreme Gradient Boosting} (XGBoost) and \textit{Random Forest}. Random Forest \cite{breiman2001random} is an algorithm that combines several decision trees and arises from the modification of the \textit{bagging} process after decorrelating the trees generated in the process. The bagging technique is meant to reduce the variance from the combination of models. During the process, the resampling technique \textit{bootstrapping} is used. With this technique new data sets are created from the extraction of samples (with repetition) of the training data set. So, in a regression problem, a model is trained with each of the new training sets and the prediction of the total model will be the combination of the output of each of the trained models. The parameters to adjust after the training stage have been clustered and evaluated with the validation data set:

\begin{itemize}
\item Number of estimators that will be built. Evaluated values: [50,100,500,1000]. Best value found: 100.

\item \textit{Maximum depth} for the trees built. Evaluated values: [10,15,20]. Best value found: 20.
\end{itemize}

The second approach considered is the Extreme Gradient Boosting (XGBoost) algorithm \cite{friedman2001greedy,chen2016xgboost} which is based on the \textit{boosting} method. This method is similar to bagging technique, but in this case the trees are built sequentially. That is, each of the trees are built using information from those previously built. In this way, ``robust'' trees are built from ``weaker'' ones. This is achieved by using the Gradient descent algorithm as the optimiser. Thus, during each iteration of the training process, the parameters of the weakest models are adjusted in order to minimise the loss function established for the problem (for instance, the root of the mean square error in the case of a regression), which will be given by the result of the model. Some parameters must be established when training this algorithm, for which we have followed the same strategy used previously. The parameters which were evaluated using the validation data set are the following:

\begin{itemize}
\item \textit{Maximum depth} for the trees built. Evaluated values: [8,9,10]. Best value found: 10.
\item \textit{Learning rate}. Evaluated values: [0.01, 0.1, 0.2, 0.7]. Best value found: 0.2.
\item \textit{Objective}: define the kind of output that will be provided by the algorithm. In our case, a probability is wanted, thus, ``binary logistic'' is chosen. 
\item{Scale\_pos\_weight}: is used to control the balance of positive and negative weights, which is useful for unbalanced classes. Introducing the ratio between the classes $N_{em}/N_{\mu}$ which is present in the preprocessed training data set helped in improving the result.
\item Maximum number of rounds when training. Selected value: 1000.
\item Early stopping rounds: applied to find the optimal number of boosting rounds. If the error is not reduced in the validation data set after a fixed number of rounds the training will stop. Selected value: 20. 
\end{itemize}

\subsection{Ensemble approach}

The previous approaches have a partial overlap in the variables used as input, however, there might be some features discovered by the CNN that tree models are not able to compute. At the same time, due to the smaller input data and the efficient optimization strategy, tree models might obtain more accurate results. With this two facts in mind, the possibility of combining both outputs in an ensemble is discussed.


Two ensembles are designed using the CNN as the main algorithm and a decision-tree based algorithm as the second algorithm, whose outputs will be obtained using the equation (\ref{eq:ensemble}). Let $P_{\mu, \mathrm{pred}} (w)$ be the probability obtained by the ensemble and let $P_{\mu, \mathrm{m_1}}$ and $P_{\mu, \mathrm{m_2}}$ be the probabilities given by the two best models found for each approach respectively. Additionally, let us propose the weight \textit{w} to adjust the influence that each algorithm will have when determining the ensemble probability, so that $P_{\mu, \mathrm{pred}} (w) \in \left[ 0,1 \right] $. 

\begin{equation} \label{eq:ensemble}
P_{\mu, \mathrm{pred}} (w) = w \cdot P_{\mu, \mathrm{m_1}} + (1-w) \cdot P_{\mu, \mathrm{m_2}}
\end{equation}

In principle, there are infinite possibilities when combining the outputs of the models depending on the weight \textit{w}. However, we can adjust and get an optimum value of \textit{w} in order to minimise the error. A possible approach to optimise the weight of each ensemble is to find the value which minimises the mean squared error after predicting the samples of the validation data set \cite{shahhosseini2019optimizing}, that is, minimizing the equation (\ref{eq:optimization}), where \textit{n} is the total number of samples in the validation data set, \textit{i} is the index of the sample, $P_{\mu, \mathrm{pred}}^{(i)}$ is the probability obtained by the ensemble and $P_{\mu}^{(i)}$ the real ``probability'' and label of that sample (1 if there is a muon in the WCD or 0 otherwise). To solve the global optimization problem \textit{Differential Evolution} algorithm from the \textit{Scipy Optimize} Python package is implemented, which is a stochastic population based method \cite{storn1997differential}.

\begin{equation} \label{eq:optimization}
\mathrm{MSE} = \frac{1}{n} \sum_{i=1}^{n} \left ( P_{\mu, \mathrm{pred}}^{(i)}(w) - P_{\mu}^{(i)} \right )^2 
\end{equation}

With the previous approach, when there is a disagreement between the models, one of them will have more weight than the other. This is translated in that, if one model predicts muon and the other does not the ensemble will predict muon unless the weight given to the other model is very small. This situation is undesirable as it is a requirement to be sure about when muons appear so two other approaches to combine the model's outputs have been considered as well:

\begin{equation} \label{eq:ensemble2}
P_{\mu, \mathrm{pred}} =  P_{\mu, \mathrm{m_1}} \cdot  P_{\mu, \mathrm{m_2}}
\end{equation}

\begin{equation} \label{eq:ensemble3}
P_{\mu, \mathrm{pred}} =  \sqrt{P_{\mu, \mathrm{m_1}} \cdot  P_{\mu, \mathrm{m_2}}}
\end{equation}

\begin{equation} \label{eq:ensemble4}
P_{\mu, \mathrm{pred}} = \frac{1}{\sqrt{2}}  \sqrt{P_{\mu, \mathrm{m_1}}^{2} +  P_{\mu, \mathrm{m_2}}^{2}}
\end{equation}

By using the product, in case of clear disagreement, the ensemble will not provide a high probability. This value will be high only when both models agree. The ensemble of equation (\ref{eq:ensemble4}) was designed with the aim of giving greater importance to the algorithm that provides the highest probability.


\section{Experiments and discussion} \label{sec:results}


This Section will show first, how relevant are the new variables proposed in Section \ref{sec:data}. Afterwards, the performance of the ensemble models is analysed. Finally, an experiment on solving the classification task using the probability is presented with the models presented in the paper.

As stated in Section \ref{sec:data}, the data set is highly imbalanced so, after making some trials with the cited strategies, the best one (evaluated using validation error) was Random oversampling with ratio $N_{\mu}/N_{\mathrm{e.m.}} = 0.5$, where $N_{\mu}$ and $N_{\mathrm{e.m.}}$ are the number of stations with muons and without muons respectively. The following experiments, unless stated explicitly, are made after applying this procedure to the training data set.

Regarding the implementation details, Python 3.7 using SciKit-learn \cite{scikit-learn}, Numpy \cite{numpyScipy} and Keras Framework \cite{keras} were used to develop the neural networks and random forest. Regarding the XGBoost, the implementation available at \cite{xgboostPython} was chosen to carry out the experiments. Due to the high volume of data (each EAS can require up to hundreds of MiB), the infrastructure used to compute was a cluster at LIP with the characteristics: Intel(R) Xeon(R) Silver 4110 CPU @ 2.10GHz processors, 46 GiB of RAM and two GPUs: NVIDIA GP102 TITAN Xp, NVIDIA TU102 GeForce RTX 2080 Ti.

For the sake of reproducibility and transparency, all the code and data are available at \url{https://github.com/borja-sg/Muon-identification-WCD}.


\subsection{Analysis of engineered variable's importance and approximation error}

To determine how important the variables are, two methods were applied. The first one is based on the ranking that the XGBoost algorithm builds during the training stage. The second is based on the approximation error achieved by each model when training with different subset of variables (given the fixed model architecture of Section \ref{sec:CNN}). 

\begin{figure}[t]
 \centering
 \includegraphics[width=0.43\textwidth]{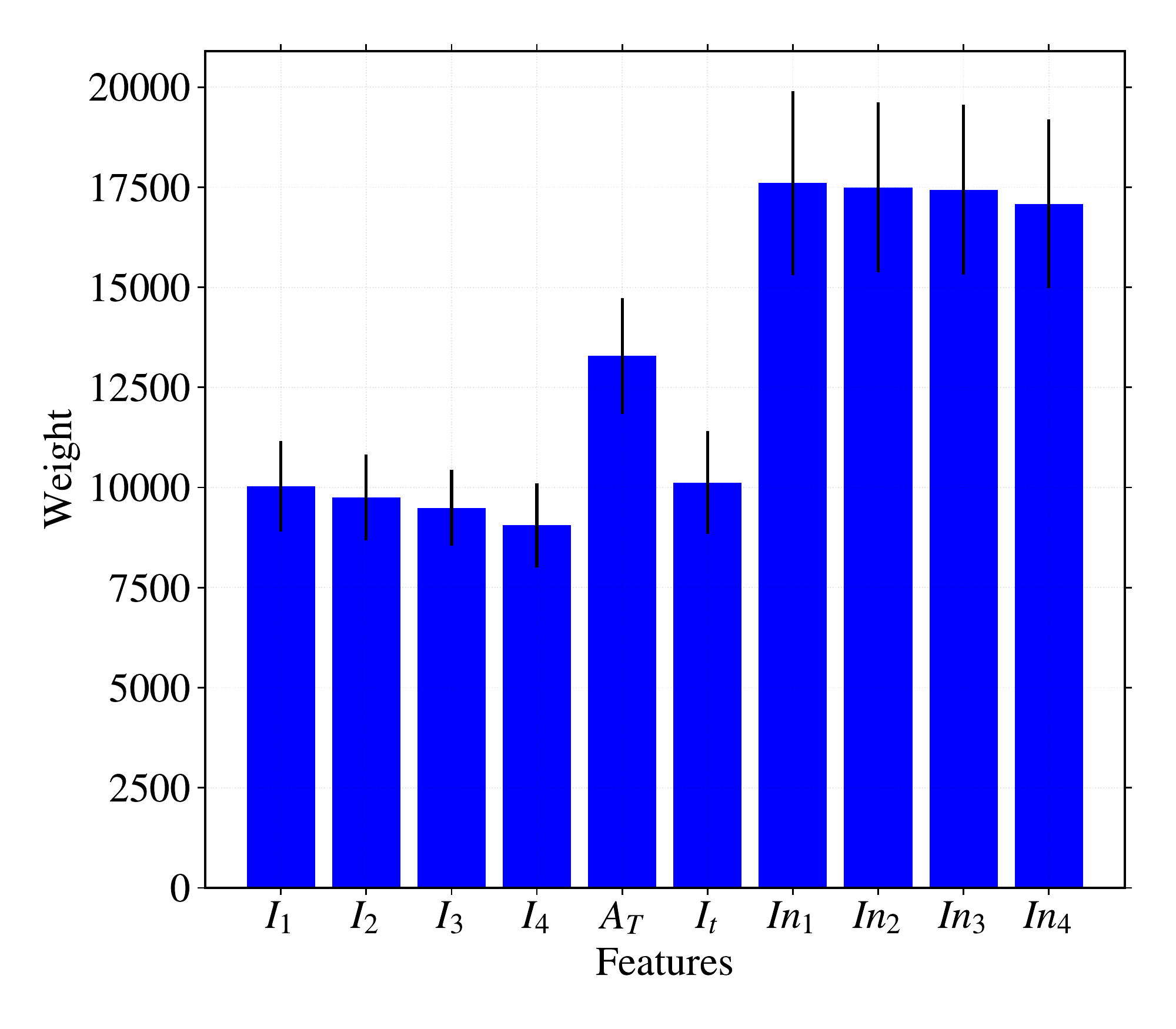}    
 \caption{Mean feature importance given by XGBoost using 5 different seeds to select train and validation data sets. The black lines correspond to the standard deviation after training with the different seeds. Importance type: \textit{weight}, which is the number of times the feature was used to split the data across all trees.}
 \label{fig:XGBoost_features_CV}
\end{figure}

Figure \ref{fig:XGBoost_features_CV} shows the weights that the XGBoost has assigned after carrying out the training. The most important variables are the integrals after the normalisation process followed by the new proposed index for the asymmetry. For the CNN approach, in Table \ref{tab:rmse-results} it is shown the mean Root Mean Squared Error (RMSE, $\sqrt{MSE}$) obtained after performing a cross validation with 5 folds. By doing this, it is guaranteed that the process is not biased by the initial shuffle to select train and validation data. As expected, the cross validation does not affect the model's learning capabilities. In this case, the combination of variables that provides the best validation results is the one that uses all the proposed variables but the asymmetry. It is fair to say that the variables proposed always improved the approximation capabilities of the CNN architecture as the CNN using only the raw normalised signals is the one that provides the worst validation error.
Table \ref{tab:rmse-results} also shows the error obtained by the tree based models. Although these models are able to learn properly the data, they tend to overfit, obtaining worse validation and test errors. Another reason to see some difference between training and validation might probably be due to the fact that the training data was balanced meanwhile validation and test are not.

\begin{table*}[h]
\centering
\begin{tabular}{|c|c|c|c|}
\hline
Input Variables                  & Train           & Validation            & Test \\ \hline 
$\vec{S},I_{i},In_{i}, I_{t}, A_{T} $  & 0.2118 (0.0049) & 0.2361 (0.0058) & 0.2341 (0.0059) \\ \hline 
$\vec{S},I_{i},In_{i}, I_{t} $  & 0.2127 (0.003) & \textbf{0.2302 (0.0059)} & \textbf{0.2285 (0.0055)}\\ \hline 
$\vec{S},In_{i}, A_{T}, I_{t} $  & 0.2196 (0.0031) & 0.2316 (0.0058) & 0.2292 (0.0060) \\ \hline %
$\vec{S},  I_{i}, A_{T}, I_{t} $ & 0.2136 (0.0034) & 0.2340 (0.0036) & 0.2318 (0.0044) \\ \hline 
$\vec{S}$                        & 0.2221 (0.0122) & 0.2484 (0.0109) & 0.2467 (0.0107)\\ \hline 
$\vec{S}, A_{T}$                 & 0.2166 (0.0076) & 0.2447 (0.0104) & 0.2438 (0.0109) \\ \hline 
$\vec{S}, I_{i}$                 & 0.2049 (0.0035) & 0.2331 (0.0070)  & 0.2312 (0.0077) \\ \hline 
$\vec{S}, In_{i}$                & 0.2194 (0.0068) & 0.2425 (0.0108) & 0.2409 (0.0106) \\ \hline 
$\vec{S}, I_{t}$                 & 0.2187 (0.0064) & 0.2424 (0.0084) & 0.2397 (0.0086) \\ \hline 
$\vec{S},I_{i}, I_{t} $          & 0.2098 (0.0032) & 0.2362 (0.0050)  & 0.2340 (0.0060)  \\ \hline 
$\vec{S},  In_{i}, I_{t} $       & 0.2182 (0.0043) & 0.2351 (0.0078)  & 0.2329 (0.0074) \\ \hline 
XGBoost       & \textbf{0.1052 (0.0121)} & 0.2340 (0.0040) & 0.2342 (0.0032) \\ \hline %
RF       & 0.2301 (0.0010) & 0.2606 (0.0013) & 0.2600 (0.0004) \\ \hline %
\end{tabular}
\caption{Mean RMSE (and standard deviation) for models trained using 5 different seeds to select train and validation data sets.}
\label{tab:rmse-results}
\end{table*}

\subsection{Comparing the ensemble approach}

From the previous experiment, we have seen that the features proposed were able to improve the probability prediction. Once the best models have been identified, it is time to see if they can combine knowledge assembling them. Table \ref{tab:ensemble-results} shows the obtained RMSE for the different strategies towards creating the ensemble's output. The two models were chosen considering the best RMSE in validation for the models using the PMT signal's (CNNs) as well as for the tree based models. It is interesting to observe how, due to the overfitting of the XGBoost, it receives much more weight than the CNN although it performs worse in validation. The weight computation is also affected by the training data preprocessing and it's important to notice how the balancing still seems necessary as it provides better validation and test errors. The most remarkable fact of this table is that, regardless of the strategy used to combine both model's outputs, it always improves the results obtained isolated. Taking a close look, it is easy to conclude that when combining probabilities, it is better to perform the product than the weighted sum. The product of both probabilities seems the most adequate as it provides significant improvement in all metrics.



\begin{table*}[]
\centering
\scriptsize
\begin{tabular}{|c|c|c|c|c|c|c|c|}
\hline
Model 1                               & Model 2  & Strategy & $w$ optimisation       & $w$  & Train & Val. & Test  \\ \hline
$\mathrm{CNN}$ ($\vec{S},I_{i},In_{i}, I_{t} $) & XGB & $w \cdot P_{\mu, \mathrm{m_1}} + (1-w) \cdot P_{\mu, \mathrm{m_2}}$  & Train (Balanced) & 0.2323  & 0.1288  & 0.2092  & 0.2102  \\ \hline
$\mathrm{CNN}$ ($\vec{S},I_{i},In_{i}, I_{t} $) & XGB & $w \cdot P_{\mu, \mathrm{m_1}} + (1-w) \cdot P_{\mu, \mathrm{m_2}}$   & Train (Original, imbalanced)          & 0.1833  & 0.1285  & 0.2131  & 0.2141  \\ \hline
$\mathrm{CNN}$ ($\vec{S},I_{i},In_{i}, I_{t} $)& XGB & $P_{\mu,m_1} \cdot P_{\mu,m_2}$  & - & -   & \textbf{0.1026}  & \textbf{0.1872}  & \textbf{0.1883}  \\ \hline
$\mathrm{CNN}$ ($\vec{S},I_{i},In_{i}, I_{t} $)& XGB & $\sqrt{P_{\mu,m_1} \cdot P_{\mu,m_2}}$   & - & -   & 0.1193  & 0.1897  & 0.1901  \\ \hline
$\mathrm{CNN}$ ($\vec{S},I_{i},In_{i}, I_{t} $)& XGB & $\frac{1}{\sqrt{2}} \cdot \sqrt{P_{\mu,m_1}^2 + P_{\mu,m_2}^2}$   & - & -   & 0.1623  & 0.2133  & 0.2126  \\ 
\hline

\end{tabular}
\caption{RMSE error for different ensemble strategies combining the best CNN with XGBoost. Notice that the approaches using the product do not require the \textit{w} optimisation parameter. }\label{tab:ensemble-results}
\end{table*}

\subsection{Tank Classification}

Once it was obtained the most accurate model to predict the probability of having a muon in the analysed WCD, this experiment will show if this prediction is reliable enough to estimate whether muon has crossed the detector or not. In this task of saying if a tank has been crossed by a muon, it has to be set a threshold to decide if the given probability is high enough. If the value is higher than the threshold, the tank is classified as positive. Otherwise, it is considered as with no muon. To determine the threshold, a loop applying cuts to the obtained probability was used with a resolution of 0.01 and choosing the best one according to a metric.



\begin{figure*}[t]
 \centering
  \subfloat[Test using CNN and XGB.]{
   \label{f:dense_array}
    \includegraphics[width=0.45\textwidth]{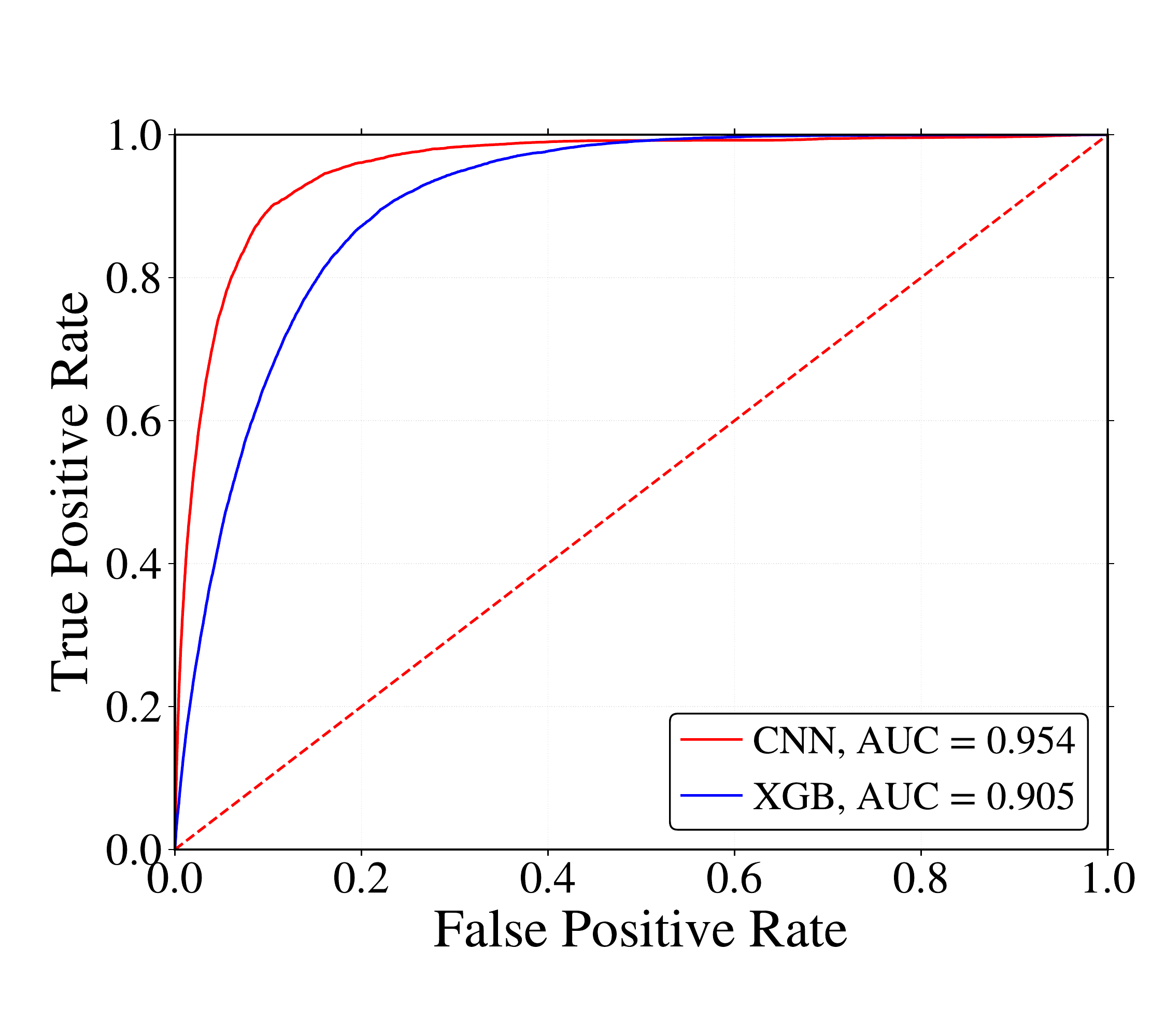}}
  \subfloat[Test using ensembles.]{
   \label{f:sparse_array}
    \includegraphics[width=0.45\textwidth]{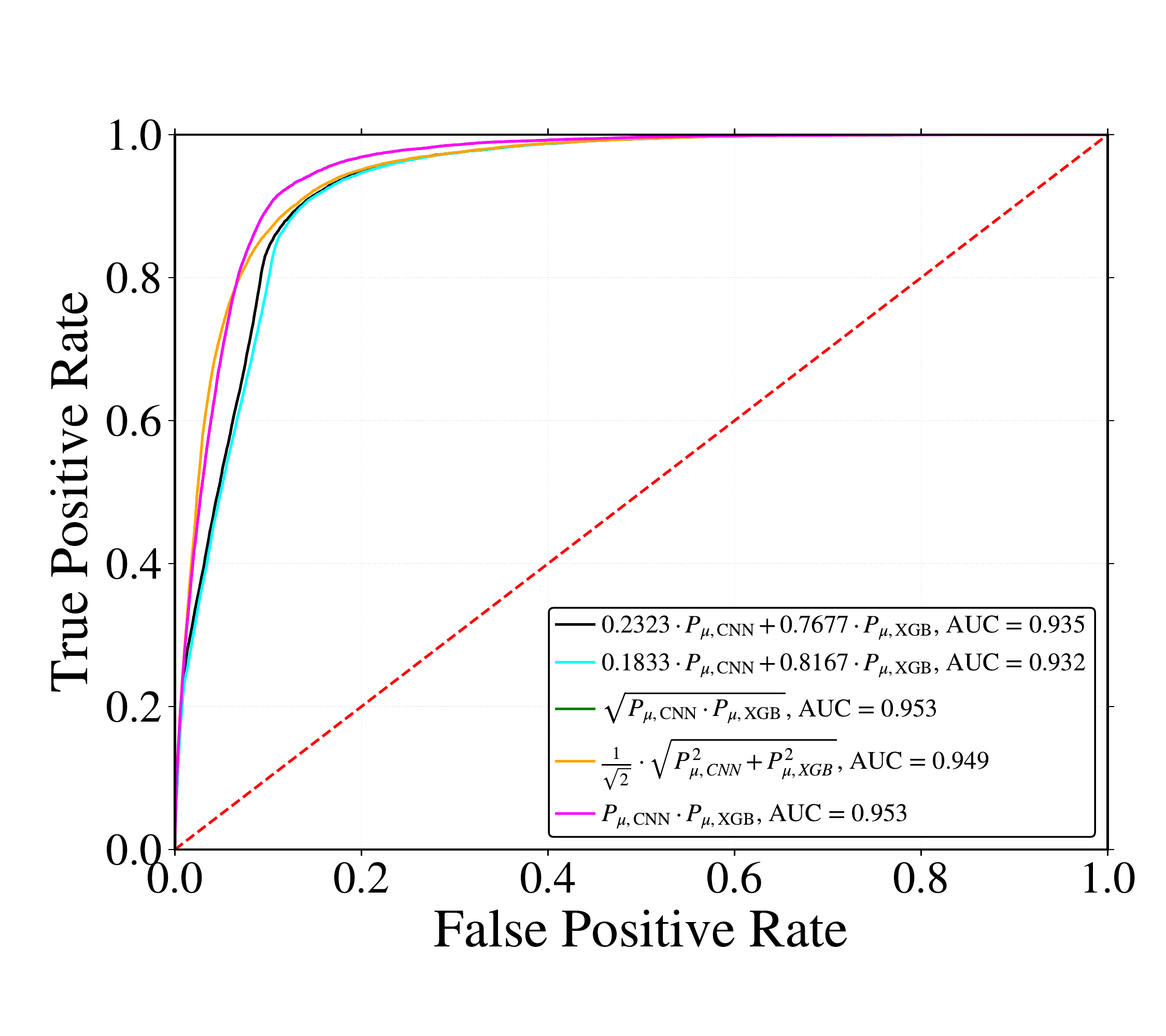}}
 \caption[Result of the ROC curves for test data set using: (a) models and (b) ensembles.]{Result of the ROC curves and AUC for test data set (a) single models (b) ensembles.}
 \label{fig:ROC_curves}
\end{figure*}

Four classification metrics have been used: Area under the \textit{Receiver Operating Characteristic} (ROC) curve, F1-score, Accuracy, and an Ad-hoc criterion. The first considered is based on the (ROC) curves of each proposed model to explore which ratio between the \textit{True Positive Rate} (TPR) and the \textit{False Positive Rate} (FPR) is suitable to determine whether there is a muon in the stations. Considering using this information in an ulterior system to classify $\gamma$/hadron EAS is necessary to ensure a low rate of false positive rate. In such a task, the ROC curves can be effective to analyse the behavior of the models when selecting different thresholds. In addition, the \textit{Area Under the Curve} (AUC) will allow us to compare the overall performance of the models. Figure \ref{fig:ROC_curves} shows the results obtained for each model for the two test data sets. According to the results obtained, there is a match between the CNN and the ensemble obtaining and excellent result for \textit{Single Muons}. 
XGBoost has a good performance but it does not achieve as good results.

For the other three metrics, Tables \ref{tab:max_results} and \ref{tab:max_results2} show the results for the three models. Table \ref{tab:max_results} shows results for the thresholds that maximise F1-score and Accuracy metrics respectively for the balanced Train data set. In this case, the ensemble outperforms in accuracy and F1-score, however, for the last metric, the CNN show a better behaviour for validation and test data sets probably due to the effects of the overfitting of the XGBoost, which shows very poor performances in both validation and tests.

The last metric is the criterion that could be used for an ulterior classification of the $\gamma$ vs hadron EAS and to identify the type of particle. In this case, it is very important to not to missclassify the $T_{\gamma}$, this is, the tanks where a muon has not passed through and, it is enough to be able to identify some muons ($T_{\mu}$). Thus, these values were computed and the cut has been set where the $T_{\gamma}$ rate is near 99.9 \%. As this implies new threshold values, the other two metrics have been computed as well to have a fair comparison of the models using different criteria. From Table \ref{tab:max_results2} the performance of both the CNN and ensemble remains quite similar although in the test sets, for the F1-score, the ensemble does an improvement in comparison with the other approaches. For the Ad-hoc criterion, we have two non-dominant solutions, this is, from the two objectives that we are aiming, $T_{\mu}$ and $T_{\gamma}$, there is no model that perform better in both at the same time. If a more accurate $T_{\mu}$ is desired, the ensemble should be chosen but a 0.2\% of error will be increased when misclassifying $T_{\gamma}$.

\begin{table*}[h!]
\scriptsize
 \centering
 \begin{tabular}{|c|c|c|c|c|c|}
 \hline
\multicolumn{6}{|l|}{\textbf{Accuracy}}                                                                                                                        \\ \hline
 Algorithm                                                                            & Threshold & Train & Train (Original, imbalanced) & Validation & Test \\ \hline %
 $\mathrm{CNN}$                                                                     & 0.27       &  94.72  &  94.04      & 92.70              &    92.73 \\ \hline %
 XGBoost                                                                              &  0.83    & 99.67   &  99.59   &   94.60   &  94.48      \\ \hline %
 $\mathrm{CNN}$ $\cdot$ XGBoost                                                     &  0.18    & 96.03  & 97.74   &  94.68    & 94.66    \\ \hline %
 
 \multicolumn{6}{|l|}{\textbf{F1-score}}                                                                                                                        \\ \hline
 Algorithm                                                                            & Threshold & Train & Train (Original, imbalanced)  & Validation & Test \\ \hline %
 $\mathrm{CNN}$                                                                     & 0.27    & 92.39  &   60.81   &  52.16  & 52.36          \\ \hline %
 XGBoost                                                                              & 0.83    &  99.51 &  95.95     &  27.86     & 27.89     \\ \hline %
 $\mathrm{CNN}$ $\cdot$ XGBoost                                                     & 0.16     & 93.92 &  78.54     &  50.14     &  50.44    \\ \hline %
 \end{tabular}
 \caption{Maximum values for two error measures when classifying the appearance of muons in the WCDs. This is the best value obtained for any possible threshold to determine the class separation. }\label{tab:max_results}
 \end{table*}

\begin{table*}[h!]
\scriptsize
 \centering
\begin{tabular}{|c|c|c|c|c|c|c|c|c|c|}
\hline
\multicolumn{10}{|l|}{\textbf{Accuracy}}                                                                                                                                                                                                      \\ \hline
Algorithm                      & Threshold                  & \multicolumn{2}{c|}{Train} & \multicolumn{2}{c|}{Train (Original)} & \multicolumn{2}{c|}{Validation} & \multicolumn{2}{c|}{Test } \\ \hline
$\mathrm{CNN}$                 & 0.96                       & \multicolumn{2}{c|}{70.70} & \multicolumn{2}{c|}{95.59}            & \multicolumn{2}{c|}{95.61}      & \multicolumn{2}{c|}{95.53}     \\ \hline
XGBoost                        & 0.94                       & \multicolumn{2}{c|}{95.69} & \multicolumn{2}{c|}{99.14}            & \multicolumn{2}{c|}{95.04}      & \multicolumn{2}{c|}{94.97}     \\ \hline
$\mathrm{CNN}$ $\cdot$ XGBoost & 0.82                       & \multicolumn{2}{c|}{87.05} & \multicolumn{2}{c|}{98.04}            & \multicolumn{2}{c|}{95.47}      & \multicolumn{2}{c|}{95.48}     \\ \hline
\multicolumn{10}{|l|}{\textbf{F1-score}}                                                                                                                                                                                                      \\ \hline
Algorithm                      & Threshold                  & \multicolumn{2}{c|}{Train} & \multicolumn{2}{c|}{Train (Original)} & \multicolumn{2}{c|}{Validation} & \multicolumn{2}{c|}{Test }  \\ \hline
$\mathrm{CNN}$                 & 0.96                       & \multicolumn{2}{c|}{22.13} & \multicolumn{2}{c|}{21.26}            & \multicolumn{2}{c|}{22.39}      & \multicolumn{2}{c|}{20.71}  \\ \hline
XGBoost                        & 0.94                       & \multicolumn{2}{c|}{93.10} & \multicolumn{2}{c|}{90.47}            & \multicolumn{2}{c|}{16.10}      & \multicolumn{2}{c|}{16.41}  \\ \hline
$\mathrm{CNN}$ $\cdot$ XGBoost & 0.82                       & \multicolumn{2}{c|}{75.94} & \multicolumn{2}{c|}{75.08}            & \multicolumn{2}{c|}{21.67}      & \multicolumn{2}{c|}{23.04}  \\ \hline
\multicolumn{10}{|l|}{\textbf{Ad-hoc criterion}}                                                                                                                                                                                              \\ \hline
\multirow{2}{*}{Algorithm}     & \multirow{2}{*}{Threshold} & \multicolumn{2}{c|}{Train} & \multicolumn{2}{c|}{Train (Original)} & \multicolumn{2}{c|}{Validation} & \multicolumn{2}{c|}{Test }  \\ \cline{3-10} 
                               &                            & $T_{\mu}$  & $T_{\gamma}$  & $T_{\mu}$        & $T_{\gamma}$       & $T_{\mu}$     & $T_{\gamma}$    & $T_{\mu}$    & $T_{\gamma}$       \\ \hline
$\mathrm{CNN}$                 & 0.96                       & 12.49      & 99.81         & 12.35            & 99.81              & 13.13         & 99.79           & 11.99        & 99.81           \\ \hline
XGBoost                        & 0.94                       & 87.25      & 99.91         & 84.11            & 99.91              & 9.86          & 99.36           & 10.14        & 99.31           \\ \hline
$\mathrm{CNN}$ $\cdot$ XGBoost & 0.82                       & 61.33      & 99.90         & 61.24            & 99.90              & 12.99         & 99.65           & 13.89        & 99.66           \\ \hline
\end{tabular}
 \caption{Error measures (accuracy and F1-score) and percentage of correct classification (Ad-hoc criterion) when classifying the appearance of muons in the WCDs. These are the best values to ensure a minimum of 
 $T_{\gamma} \sim 99.9 \%$ using the balanced Train data set.}\label{tab:max_results2}
 \end{table*}

After comparing the different models using several metrics, it is quite obvious that the best approach in general is the one that receives the input the traces and the engineered variables, this is, $\mathrm{CNN}$. From this fact, it is possible to devise that the temporal and spatial components of the signals are crucial when trying to discriminate the presence of muons. One reason for this might be the fact that the traces are seeing 30 ns from the signal that includes the direct light and the first reflection, meanwhile the integral is computed considering the first 10 ns to avoid too much interference with the electromagnetic component. Thus, a tank that maximises reflections should be considered when designing the new observatory. Another interesting element to discuss is inability of models to learn the engineered variables. This makes sense as the engineered variables are almost impossible to be obtained just by performing convolutions but it is a call of attention to DL practitioners so they do not rely uniquely on the model but they have to put special attention to the problem domain and the expert knowledge available.


\section{Conclusions} \label{sec:conclusions}
The muon identification is a hot topic in cosmic ray research due to its implications in $\gamma$-ray observation, multimessenger studies and Ultra High Energy Physics. Within this context, a new observatory based on indirect methods, like Water Cherenkov Detectors, is being studied. This paper has dealt with the problem of determining the probability of muon presence in the signal recorded by a water Cherenkov detector prototype design. To do so, several machine learning techniques have been compared and new engineered variables have been proposed. The results have shown that the models that behave the best considering the characteristic of the problem are the ones that process the complete signal and that introduce the variables engineered by researchers in Physics and Computer Science by means of a combination of Convolutional Layers with Dense layers. Although the improvement is small, it is magnified when using an ensemble of models combining CNNs and XGBoost. Therefore, the possibilities that this new methodology open to identify Extensive Air Showers with $\gamma$-rays are wide open and allow evaluating the different detector designs before deploying the future observatory. Future studies will concentrate in considering other convolutions for the CNN and to consider methods that do not suffer from overfiting as much as the XGBoost does with these data sets.

\section*{Acknowledgements}
We would like to thank to A. Bueno for all the support and useful discussions during the development of this work.
The authors thank also for the financial support by OE - Portugal, FCT, I. P., under project PTDC/FIS-PAR/29158/2017.
R.~C.\ is grateful for the financial support by OE - Portugal, FCT, I. P., under DL57/2016/cP1330 /cT0002. 
B.S.G. is grateful for the financial support by grant LIP/BI - 14/2020, under project IC\&DT, POCI-01-0145-FEDER-029158. 

\bibliography{references}

\end{document}